\documentclass{article}
\usepackage[utf8]{inputenc}
\usepackage{hyperref}

\title{Automatic Historical Stock Price Dataset Generation Using Python}

\author {
    Arunima Mandal, Yuanhang Shao, Xiuwen Liu \textsuperscript{\rm 1}\\
    \textsuperscript{\rm 1} Florida State University\\
    amandal@cs.fsu.edu, shao@cs.fsu.edu, liux@cs.fsu.edu
}

\date{}

\usepackage{graphicx}
\usepackage{listings,newtxtt}
\usepackage{times}

\begin{document}

\maketitle
\section*{ABSTRACT}
With the dynamic political and economic environments, the ever-changing stock markets generate large amounts of data daily. Acquiring up-to-date data is crucial to enhancing predictive precision in stock price behavior studies. However, preparing the dataset manually can be challenging and time-demanding. The stock market analysis usually revolves around specific indices such as S\&P500, Nasdaq, Dow Jones, the New York Stock Exchange (NYSE), etc. It is necessary to analyze all the companies of any particular index. While raw data are accessible from diverse financial websites, these resources are tailored for individual company data retrieval and there is a big gap between what is available and what is needed to generate large datasets. Python emerges as a valuable tool for comprehensively collecting all constituent stocks within a given index. While certain online sources offer code snippets for limited dataset generation, a comprehensive and unified script is yet to be developed and publicly available. Therefore, we present a comprehensive and consolidated code resource that facilitates the extraction of updated datasets for any particular time period and for any specific stock market index and closes the gap. The code is available at \url{https://urldefense.com/v3/__https://github.com/amp1590/automatic_stock_data_collection__;!!PhOWcWs!xaSIXxnrsG-KdI-EGOIXYISvDgsHGwV0b7smpOYFDPQlRZJjhSBYMospmGAsigtDHQkg0meusbNYsMY$ }.

\section{INTRODUCTION}
An immense amount of financial data is produced daily from the stock market and is easy to obtain from various websites like Yahoo Finance, Google Finance, Zacks, TipRanks, etc. However, finding a comprehensive and well-organized dataset covering a wide range of companies can be quite a repetitive and daunting task. Numerous valuable resources exist online offering guidance on dataset acquisition\cite{ding2015deep}. However, a majority of them do not provide complete and working programs, covering all requisite steps — such as extracting the list of companies from a specific stock market index, refining the list, storing data in a designated directory, downloading the most valuable information, and saving individual company data in .csv format.

A novel Python package called ``stocksymbol'' \footnote{Available at \url{https://urldefense.com/v3/__https://pypi.org/project/stocksymbol/__;!!PhOWcWs!xaSIXxnrsG-KdI-EGOIXYISvDgsHGwV0b7smpOYFDPQlRZJjhSBYMospmGAsigtDHQkg0meuZlg9Elg$ }} claims to ease the process of collecting stock symbols from different exchanges. This package collects data from the Yahoo site rather than from the wiki or the respective original websites of different stock market indices. This makes the stock symbols or tickers collecting part much easier in a concise way but not applicable to all the stock market indexes. 
For example, S\&P500 generates perfectly but not Nasdaq, which generates only 1500 companies whereas the stock symbols of Nasdaq collected from datahub.io contain about 3000 stock symbols. Therefore, mining the stock symbols from specific URLs is preferred instead of relying on the ``stocksymbol'' package.

In order to facilitate the acquisition of historical datasets, we design and implement a comprehensive Python script. This script necessitates minimal adjustments within the code, primarily concerning the desired interval (daily/weekly/monthly) and the date range. 

Our script supports various data sources such as Yahoo Finance, Google Finance, Quandl, Tiingo, IEX Cloud, and more. Specifically, we leverage the extensively used Yahoo Finance data source, which conveniently requires no registration. In this paper, we provide code examples for gathering the S\&P 500 and Nasdaq datasets. These two datasets stand as the most commonly employed resources in the USA for conducting stock market analyses.
 
\section{ABOUT HISTORICAL DATASET}
The Yahoo Finance API contains a vast dataset of historical stock prices and other financial information for publicly traded companies from major indices such as the S\&P 500 and Nasdaq, as well as less well-known, smaller businesses. The data provides a range of financial parameters as individual columns that are saved in a CSV format file. Each file is named after its respective stock symbols and will contain the following columns:
\begin{itemize}

\item Date: The date of the trading day.
\item Open: The opening price of the stock for that day.
\item High: The highest price the stock reached during that day.
\item Low: The lowest price the stock reached during that day.
\item Close: The closing price of the stock for that day.
\item Volume: The number of shares of that stock that were traded during that day.

\end{itemize}

The ``Adj\_close'' column in the original downloaded dataset of Yahoo Finance, refers to the adjusted closing price taking dividends, stock splits, and new stock offerings into account. The adjusted closing price is a superior indicator of stock value as it begins where the closing price ends. However, this column is not included explicitly because all the presented prices are adjusted in the code for potential corporate actions.

Our dataset collects these files of all the companies from any particular stock index. These .csv files do not contain the name of the company as a column, but we need that in our dataset. So we also add another column - ``Name'' as the last column which contains the name of the company or ticker.

\section{COLLECTING DATA USING YFINANCE PACKAGE}
The stock data can be downloaded from different packages such as Yahoo Finance, Quandl, and Alpha Vantage. We use the ``yfinance'' module which has gained immense popularity as a Python library. It is highly compatible with Python and can serve as an extension to ``pandas\_datareader'' or function independently. Initially named ``fix\_yahoo\_finance'', it evolved into its own module, though it is not endorsed by Yahoo as an official tool. Embraced by numerous users, it offers diverse applications, notably for acquiring stock and cryptocurrency prices.

\subsection{COLLECTING STOCK SYMBOLS FOR DIFFERENT INDEXES/COMPANY LIST FROM AN INDEX}
Different stock market indexes have different sources to get the company list and different ways to process them. Therefore, except this, all the other parts of the code would be the same for any stock market index.
\subsubsection{S\&P 500 dataset}
The S\&P500 index contains 505 companies. We collect the S\&P500 dataset from the wikipedia\cite{wikipedia-contributors-2023}. Even though Wikipedia is not an official source, we will mine that source since it is very easy to implement with Python.
Here are the Python packages that were used for S\&P500:
\begin{lstlisting}[language=Python]
    import yfinance as yf
    import pandas as pd
    import shutil
    import time
    import datetime
    import csv
    import os
\end{lstlisting}
The Wiki page contains a list of companies in an HTML table and Pandas has the option to convert HTML tables from URLs directly to DataFrame. There are 2 tables on that page, where the list of companies is obtained from the ``Symbol'' column of the first table. The datahub.io webpage also contains the list of S\&P500 companies but they also fetch that from the wiki page as mentioned in their documentation.

\begin{lstlisting}[language=Python]
    sp500url='https://urldefense.com/v3/__https://en.wikipedia.org/wiki/List_of_S*26P_500_companies__;JQ!!PhOWcWs!xaSIXxnrsG-KdI-EGOIXYISvDgsHGwV0b7smpOYFDPQlRZJjhSBYMospmGAsigtDHQkg0meuOhAt5PI$ '
    data_table=pd.read_html(sp500url)
    ticker_list = data_table[0]['Symbol'].tolist()
\end{lstlisting}
After gathering the list of companies, we need to clean the data because there are two companies in the Wikipedia listing whose symbol does not match the symbol in the Yahoo finance listing from where we collected the data. We replace those two companies' symbols with the appropriate symbols by the following code.

\begin{lstlisting}[language=Python]
    for i in range(len(ticker_list)):
        if ticker_list[i] == 'BRK.B':
            ticker_list[i]='BRK-B'
        elif ticker_list[i] == 'BF.B':
            ticker_list[i]='BF-B'  
\end{lstlisting}
The list at present contains 503 company symbols. It may differ over time.

\subsubsection{Nasdaq dataset}
The Nasdaq index contains more than 3,300 companies. Here, we collect the Nasdaq company list from the datahub.io site which lists most of the Nasdaq-listed companies(about 2967 companies). They fetch the list of companies from the Nasdaq's official webpage. If the Nasdaq-100 companies are needed, they can be obtained simply from the Wikipedia page (\url{https://urldefense.com/v3/__https://en.wikipedia.org/wiki/Nasdaq-100__;!!PhOWcWs!xaSIXxnrsG-KdI-EGOIXYISvDgsHGwV0b7smpOYFDPQlRZJjhSBYMospmGAsigtDHQkg0meurN7sOHA$ }) same as S\&P500. The code for scraping Nasdaq-100 company list is as follows:

\begin{lstlisting}[language=Python]
    nasdaq100url='https://urldefense.com/v3/__https://en.wikipedia.org/wiki/Nasdaq-100__;!!PhOWcWs!xaSIXxnrsG-KdI-EGOIXYISvDgsHGwV0b7smpOYFDPQlRZJjhSBYMospmGAsigtDHQkg0meurN7sOHA$ '
    data_table=pd.read_html(nasdaq100url)
    ticker_list = data_table[4]['Ticker'].tolist()
\end{lstlisting}
Sometimes, the number of tables on the URL page may vary so to get the appropriate table, different indexes might be tried if data\_table[4] does not work.
To fetch all Nasdaq companies listed in datahub.io, two more Python packages are required as follows along with the packages for S\&P500:

\begin{lstlisting}[language=Python]
    import requests
    import io
\end{lstlisting}
Here, fetching the dataset from the datahub URL is a bit different from the S\&P500 one, in which no cleaning step is required as the company symbols or tickers listed on this datahub source match exactly the ones on the Yahoo finance site. 
\begin{lstlisting}[language=Python]
    nasdaq_url="https://urldefense.com/v3/__https://pkgstore.datahub.io/core/nasdaq-listings/nasdaq-listed_csv/data/7665719fb51081ba0bd834fde71ce822/nasdaq-listed_csv.csv__;!!PhOWcWs!xaSIXxnrsG-KdI-EGOIXYISvDgsHGwV0b7smpOYFDPQlRZJjhSBYMospmGAsigtDHQkg0meus7Bim5A$ "
    s = requests.get(nasdaq_url).content
    companies = pd.read_csv(io.StringIO(s.decode('utf-8')))
    ticker_list = companies['Symbol'].tolist()
\end{lstlisting}
The datahub URL consists of almost all the Nasdaq companies, 2967 companies at present.

\subsection{COLLECTING QUERY DATE RANGE AND INTERVAL}

The date interval is set to 1 day by default and can be externally specified with values such as 1m, 5m, 15m, 30m, 60m, 1h, 1d, 1wk, 1mo, and more, where \emph{m} refers to minute, \emph{h} refers to hour, \emph{d} refers to day, \emph{wk} refers to week and \emph{mo} refers to month. We specify the start date and end date of the target timeline for the dataset query. The date format would be ``yyyy-mm-dd''. In our code, we collect data for 5 years from $1^{st}$ Jan 2018 to $31^{st}$ Dec 2022. Both interval and date range are required as input by the user for their requested data.

\subsection{SAVING DATA}
To save the data, we create a directory at the current working directory where this Python script is being run. Here, shutil.rmtree() is used to remove the directory if it already exists.

\begin{lstlisting}[language=Python]
directory = "data_sp500stocks"
absolute_path = os.path.dirname(__file__)
relative_path = os.path.join(absolute_path, directory)
if os.path.exists(relative_path):
    shutil.rmtree(relative_path)
os.mkdir(relative_path)
\end{lstlisting}

\subsection{COLLECTING HISTORICAL DATA OF ALL THE COMPANIES}
We use the download() function of the yfinance package to simply download the data and put it inside a loop to collect all company's data. 
\begin{lstlisting}[language=Python]
    for ticker in ticker_list:
        df = yf.download(tickers=ticker,start=start_date,end=end_date, interval=interval, auto_adjust=True, rounding=True, progress=False)
\end{lstlisting}
Here, the parameter ``rounding= True'' ensures the values up to two decimal places. We can set the rounding parameter as ``False'' if we want the precise values to be used in the dataset. One crucial step here is to provide an additional argument which is ``auto-adjust=True'' so that all the OHLC (Open/High/Low/Close) prices are adjusted for potential corporate actions such as splits. Moreover, by default, the parameter ``progress'' is set to True in the download() function of yfinance. One can set it to False i.e. ``progress=False'' if the data is small but it should be set to True for a high volume of data.

Now, the Date is the index of this dataset and not a column after converting the data in a DataFrame. To perform any data analysis on this data, we need to convert this index into a column. 
\begin{lstlisting}[language=Python]
df.reset_index(inplace=True)
df = df.rename(columns = {'index':'Date'})
\end{lstlisting}
After adding another column - ``Name'', we extract the DataFrame object to a CSV file and save each file by its respective ticker name in the target directory. The data is not generated for those companies that were nonexistent during the whole period of the requested date range. Due to their data is not available for that particular date range, this will cause an error while running the code. To manage this issue, we check whether the length of the obtained DataFrame is $0$ or not. We also keep a counting variable to track the number of companies retrieved for a particular timeline. This information is kept only for the user's knowledge.

\begin{lstlisting}[language=Python]
    ignored_stocks=0
    for ticker in ticker_list:
        df = yf.download(tickers=ticker,start=start_date,end=end_date, interval=interval, auto_adjust=True, rounding=True, progress=False)
        if len(df)==0:
            ignored_stocks=ignored_stocks+1
            print("The company ",ticker," didn't exist in this entire time period")
            continue
        df.reset_index(inplace=True)
        df = df.rename(columns = {'index':'Date'})
        df["Name"]=ticker
        df.to_csv(relative_path+"/"+ticker+".csv", index=False)
    print("Total no. of companies whose data has been collected: ", len(ticker_list)-ignored_stocks)
\end{lstlisting}

\section{COLLECTING DATA WITHOUT USING ANY FINANCE PACKAGE}
There is another way of achieving the same goal without using the yfinance package of Python. There would be a slight change in code in this case and a few extra lines of code need to be added. Except for downloading the data from Yahoo Finance's website and setting the date range, the rest of the code is the same.  We download it from a URL associated with the Download option on the Yahoo Finance website. 
In this process, we use a query string located in the ``download'' option when users attempt to download the data directly from the Yahoo Finance website by specifying the interval and the date range for any particular company. The hyperlink of the ``download'' button directs to a .csv file. We can use Panda's read\_csv() function to read this query string and convert it into a DataFrame. 

\begin{lstlisting}[language=Python]
query_string = f'https://urldefense.com/v3/__https://query1.finance.yahoo.com/v7/finance/download/*7Bticker*7D?period1=*7Bstart_date*7D&period2=*7Bend_date*7D&interval=*7Binterval*7D&events=history&includeAdjustedClose=true__;JSUlJSUlJSU!!PhOWcWs!xaSIXxnrsG-KdI-EGOIXYISvDgsHGwV0b7smpOYFDPQlRZJjhSBYMospmGAsigtDHQkg0meuNVWSHUE$ '
\end{lstlisting}
The URL that is used to collect the data consists of several pieces of information as follows and need to be processed first:
\begin{itemize}
    \item Stock symbol or ticker
    \item Date range
    \item Interval (Daily/weekly/monthly data)
\end{itemize}
For setting the timeline i.e. start date and end date, we need to convert the target date range into integer format as required for the query string(the link address of the ``download'') from where we fetch the historical data.

\begin{lstlisting}[language=Python]
    start_date = int(time.mktime(datetime.datetime(2020, 12, 31, 23, 59).timetuple()))
    end_date = int(time.mktime(datetime.datetime(2021, 6, 29, 23, 59).timetuple()))
\end{lstlisting}
And since the adjusted closing price should be considered instead of the last close price, we replace the Close column with the ``Adj Close'' column.

\begin{lstlisting}[language=Python]
    df=df.drop(columns=['Close'])
    df.rename(columns = {'Adj Close':'Close'}, inplace = True)
\end{lstlisting}
Moreover, the rounding of the column values up to two decimal places needs to be done manually in this loop. 

\begin{lstlisting}[language=Python]
    for column in df.columns:
        if column != 'Date' and column!='Name' and df[column].dtype == 'float64':
            df[column] = df[column].round(2)
\end{lstlisting}

\section{CONCLUSION}
Compiling update-to-date stock price datasets is a repetitive and time-consuming task. While code snippets are available for limited data generation, we presented a most straightforward and easy-to-use program to fetch datasets through the Python ``yfinance'' module, bypassing the need to use any specialized package to construct the stock datasets. In addition, we developed another program that does not rely on the ``yfinance'' module. 
We have tested both programs and made them publicly available. We hope the programs will help the stock prediction research community by reducing the burden of updating and generating datasets.

\bibliographystyle{plain}
\bibliography{refs}

\begin{thebibliography}{1}

\bibitem{wikipedia-contributors-2023}
Wikipedia contributors.
\newblock {List of S\&P 500 Companies}.
\newblock {\em Wikipedia}, 8 2023.

\bibitem{ding2015deep}
Xiao Ding, Yue Zhang, Ting Liu, and Junwen Duan.
\newblock Deep learning for event-driven stock prediction.
\newblock In {\em Twenty-fourth international joint conference on artificial
  intelligence}, 2015.

\end{thebibliography}


\end{document}